# A Formal Verification Methodology for Checking Data Integrity

Yasushi Umezawa, Takeshi Shimizu
*Fujitsu Laboratories of America, Inc., Sunnyvale, CA, USA*
yasushi.umezawa@us.fujitsu.com, takeshi.shimizu@us.fujitsu.com

**Abstract**

*Formal verification techniques have been playing an important role in pre-silicon validation processes. One of the most important points considered in performing formal verification is to define good verification scopes; we should define clearly what to be verified formally upon designs under tests. We considered the following three practical requirements when we defined the scope of formal verification. They are (a) hard to verify (b) small to handle, and (c) easy to understand.*

*Our novel approach is to break down generic properties for system into stereotype properties in block level and to define requirements for Verifiable RTL. Consequently, each designer instead of verification experts can describe properties of the design easily, and formal model checking can be applied systematically and thoroughly to all the leaf modules.*

*During the development of a component chip for server platforms, we focused on RAS (Reliability, Availability, and Serviceability) features and described more than 2000 properties in PSL. As a result of the formal verification, we found several critical logic bugs in a short time with limited resources, and successfully verified all of them. This paper presents a study of the functional verification methodology.*

## 1. Introduction

Logic verification has become more important but difficult to complete with increasing size and complexity of system on chip (SoC) designs. Powerful formal verification methods have been playing an important role in pre-silicon validation processes. On the other hand, it is still not realistic to verify all the functions of a SoC designs by formal verification methods because it requires lots of effort to describe design properties strictly in formal language. Even if such models are successfully developed, model checking for complex designs may be beyond the power of available tools and computing resource, resulting in fail. Therefore, hybrid verification methodologies using both formal verification and logic simulation have been discussed and proposed [1][2][3][4][5][6][7][8].

One of the most important and difficult points considered in hybrid verification methodology is to define a good verification scope of formal verification; i.e. what to be verified with formal verification techniques and what to be left for conventional validation with logic simulation. Needless to say, the goal of the project is to verify the design and minimize the verification effort, but not to apply formal verification methodology. Thus, the following three practical requirements should be considered when the formal verification criteria are defined.

- Properties defined in the formal verification scope are hard to validate thoroughly in conventional logic simulation. If such properties are solved formally, it significantly increases the validation coverage.
- The problem size of the properties is suitable for available formal verification tools. Description of the properties in formal language should be simple, and it should not take much time in model checking, so that we could insist superiority and efficiency against conventional logic simulation methods.
- The purpose and the methodology of formal verification should be well documented and shared by the design and verification team. It is important to deploy the formal verification systematically in the design flow.

In a development project of a component chip for server platforms, we considered to apply formal verification techniques from the viewpoint described above. The component chip has strong requirements for RAS features and they should be verified thoroughly before tape out.

In the following sections, the requirements and the overview of the target design are mentioned. In section three, design properties that are derived from the requirements are described in detail. In section four the design flow for verification is shown. Section five contains the result and analysis of the formal verification activity. Section six is our conclusions.

## 2. Target Design Overview

We have been developing a component chip for servers. Table 1 shows the overview of the chip implementation.

The chip requires so-called 'main-frame class' reliability so that it supports enhanced RAS features to detect soft errors in the chip and permanent failures of a component. All the internal data paths, registers, state machines, and counters are protected by parity bits. When a parity error is detected, the error must be logged and reported according to the severity of the error. The requirements in implementation level are summarized as follows.

- Parity protection in data paths.
- Parity protection in key control structures, such as FSM and counters.
- Illegal state detection for FSM and counters.

As a result of initial investigation, we found more than 1300 checkpoints for data integrity derived from the chip specification. It is not realistic to verify this huge number of checkpoints exhaustively and efficiently by logic simulation, simply because it will consume too much time.

This is the initial motivation for us to set the scope of formal verification as data integrity checking. The advantages are summarized as the following. First, the description of the properties for integrity check is easy and simple because the properties are broken down into module level. Thus, each designer can easily describe the properties, and model checking can be applied locally at each leaf module, as described in the later section. Second, the properties are data-centric and exhaustive search is needed. Formal model checking is right to use for the purpose. Also, it is possible to model the property in a simple and comprehensive description.

For the above two reasons, we could expect significant increase in productivity with formal model checking.

**Table 1. Chip implementation**

| Item | Implementation |
|---|---|
| Chip die size | 12.8 x 12.5 mm$^2$ |
| Technology | 0.11 um CMOS ASIC |
| Logic size | 3.5M gates |
| Core frequency | 250MHz |

## 3. Properties in Leaf Module Level

We reached the following three properties for data integrity verification in leaf module level after breaking down the system level properties in the previous section.

Note that we used user-written properties, and automatic assertion extraction was not performed.

- Ability of error detection.
- Soundness of internal states.
- Output data integrity.

Figure 1 shows abstraction of each module. In the following sections, each property above is explained based on the abstraction. The state A in the figure is for internal FSM and protected by odd parity. EC/ED can inject errors arbitrary into the state A. The state B is for data path and protected by odd parity.

Three properties should be described and verified against all non-structured modules or leaf modules. A leaf module should be small enough for formal verification tools so that Divide-and-Conquer approach is the key to success for verification goals. A leaf module can be excluded if it has no internal state and no data paths with parity protection. Other important properties can be verified with formal verification as well, but that is not mandatory criteria.

Note that we picked up Property Specification Language (PSL) [9] for property description because it is in widespread use as an industry standard and supported by several formal verification tools.

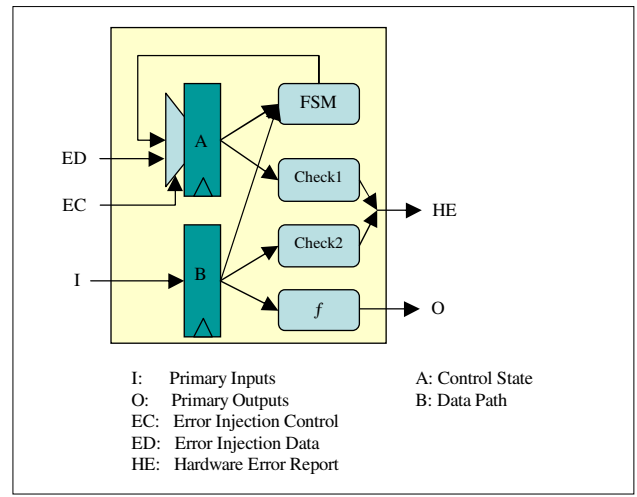

I: Primary Inputs  A: Control State
O: Primary Outputs  B: Data Path
EC: Error Injection Control
ED: Error Injection Data
HE: Hardware Error Report

**Figure 1. Typical leaf module**

### 3.1. Ability of Error Detection

The first property is for error detection. The design property is to check if all the illegal values are detected and reported at each integrity check point. Figure 2 shows a PSL code for this property, referring to the typical leaf module in Figure 1. In the PSL code, Check1 and Check2 are corresponding to pCheck1 and pCheck2 respectively. Here we have two data integrity check points.

- Check1: HE (Hardware Error Report) is true in the next cycle when EC/ED is driven and the injected value through ED is illegal. EC should be defined for each FSM and counter, but ED can be shared between FSM and counters.
- Check2: HE is true in the next cycle when I is illegal.

This property means that Check1 and Check2 in Figure 1 should fire when an error is injected.

```
vunit M_edetect (M) {            // check error detection ability
  property pCheck1 = always ((EC & ~(^ED)) -> next HE);
                                 // ED should be odd parity
  assert   pCheck1;              // -- check it formally!
  property pCheck2 = always ( ~(^I) -> next HE);
                                 // I should be odd parity
  assert   pCheck2;              // -- check it formally!
}
```

**Figure 2. PSL code for checking ability of error detection**

### 3.2. Soundness of Internal States

The second property is for soundness of internal states. The design property is to check if the integrity of internal states holds as long as the integrity of primary inputs holds.

- HE should not be asserted when no error is injected and integrity of I holds.

This property means that Check1 and Check2 in Figure 1 should not fire in normal operation. Figure 3 shows a PSL code for the property. Two properties (pIntegrityI and pNoErrInjection) are assumed, and one property (pNoError) is verified.

```
vunit M_soundness (M) {          // soundness check
  property pIntegrityI     = always ( ^I );
                                 // I should be odd parity
  assume   pIntegrityI;          // -- assumption for I
  property pNoErrInjection = always ( ~EC );
                                 // Error injection is disabled
  assume   pNoErrInjection;      // -- assumption for EC
  property pNoError        = never  ( HE );
                                 // then no error is reported
  assert   pNoError;             // -- check it formally!
}
```

**Figure 3. PSL code for checking soundness of internal states**

### 3.3. Output Data Integrity

The third property is for output data integrity. The design property is to check if the integrity of primary outputs holds as long as the integrity of primary inputs holds.

- The integrity of O should hold when no error is injected and integrity of I holds.

This property means that primary output O in Figure 1 should not have parity error in normal operation. Figure 4 shows a PSL code for the property. Two properties (pIntegrityI and pNoErrInjection) are assumed, and one property (pIntegrityO) is verified.

```
vunit M_integrity (M) {          // integrity check
  property pIntegrityI     = always ( ^I );
                                 // I should be odd parity
  assume   pIntegrityI;          // -- assumption for I
  property pNoErrInjection = always ( ~EC );
                                 // Error injection is disabled.
  assume   pNoErrInjection;      // -- assumption for EC
  property pIntegrityO     = always ( ^O );
                                 // then integrity of O holds
  assert   pIntegrityO;          // -- check it formally!
}
```

**Figure 4. PSL code for checking output data integrity**

## 4. Design Flow for Verification

This section describes the design flow of formal verification we adopted. Figure 5 shows the overall design flow at the front-end side. The logic designers are in charge of releasing *Verifiable* RTL, which is explained in section 4.1, test scenarios for functional verification, and properties or specification of data integrity. A dedicated engineer was assigned for formal verification. The verification engineer creates PSL codes based on the specification of data integrity, performs model check, and makes feedback the results to the logic designers.

### 4.1. Tasks of Logic Designers

First of all, the logic designers need to release the Verifiable RTL code. The Verifiable RTL code should satisfy the following requirements.

- Simple error injection method against every integrity check point is well-defined through primary input ports.
- Error injection should be controlled independently per entity for integrity checking.

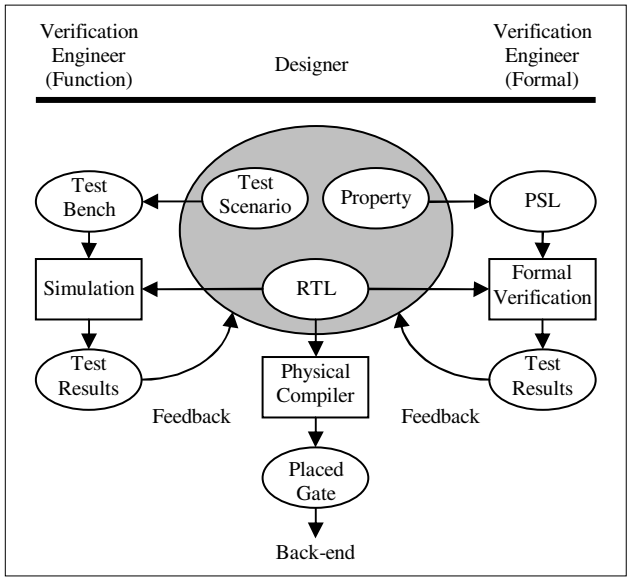

**Figure 5. Design flow**

RTL can be Verifiable by adding one line of code per such entity. The error injection ports should be tied to zero in the upper layer module because it is not used in real silicon. Figure 6 shows an example of such Verifiable RTL code in verilog.

The second task of logic designers for formal verification is to release the specification of data integrity. It can be either in text format or described as PSL codes.

### 4.2. Tasks of Verification Engineer

Three tasks are assigned for formal verification engineer.

First task is to develop PSL codes if the integrity specification is written in text format. For portability of the codes, we tried not to use tool dependent features in the process.

The second task is to perform model checking and debugging. We also consider the size of the properties in this process. When a property turned out beyond the power of available tools, for example time out happens during execution; the property can be divided for making small corns. Figure 7 shows an example of how to divide a property. The original property in Figure 7 (1) is for the output data integrity of Data D; i.e., the integrity of Data D should hold as long as the integrity of Data A, Data B, and Data C holds. The property is manually divided as shown in Figure 7 (2). The followings are the details of the divided properties.

- the integrity of Data A' should hold as long as the integrity of Data A holds
- the integrity of Data B' should hold as long as the integrity of Data B holds
- the integrity of Data C' should hold as long as the integrity of Data C holds
- the integrity of Data D should hold as long as the integrity of Data A', Data B' and Data C' holds

The last task is to feedback the results of property checking to logic designers for bug fix and verifiability.

```
module A (...);              // wrapper module
...
B B_in_A (
   ...
   .I_ERR_INJ_C ( 2'b00 ),
   .I_ERR_INJ_D ( 4'b0000 ),
 );
endmodule

module B (...);              // leaf module
input [1:0] I_ERR_INJ_C;
input [3:0] I_ERR_INJ_D;
...
reg [3:0] cs, ns;            // for FSM (including parity)
reg [3:0] cnt, cnt_n;        // for counters
...
always @(posedge CK or posedge RESET)
  if( RESET )                cs <= 4'b1_000;
  else if( I_ERR_INJ_C[0] )  cs <= I_ERR_INJ_D;
  else                       cs <= ns;
always @(posedge CK or posedge RESET)
  if( RESET )                cnt <= 4'b1_000;
  else if( I_ERR_INJ_C[1] )  cnt <= I_ERR_INJ_D;
  else                       cnt <= cnt_next;
endmodule
```

**Figure 6. An example of Verifiable RTL code**

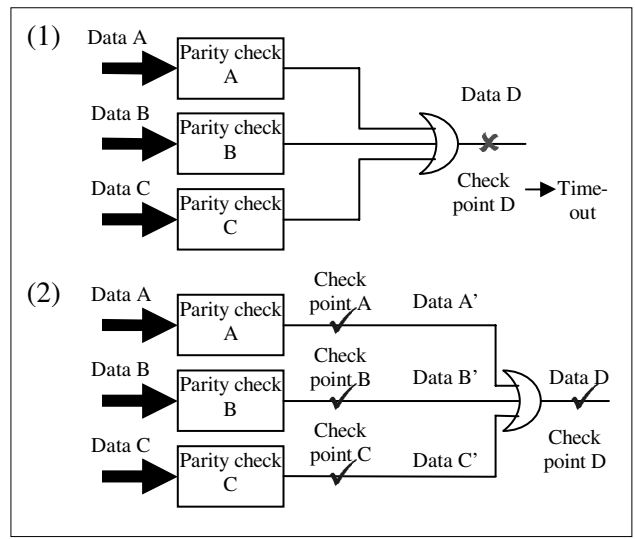

**Figure 7. Partitioning a property for "Divide-and Conquer" approach**

## 6. Results

We used a commercial formal verification tool and an in-house formal verification engine. The commercial formal verification tool is capable of handling PSL properties and verilog RTL codes, equipped with various formal solver algorithms.

The in-house formal verification engine is a powerful solver for properties with UMC (unbounded model checking) based on POBDDs (Partitioned reduced Ordered Binary Decision Diagrams) and its related algorithms as well as combined forward and backward traversal for OBDD-based invariant checking. [10]

### 6.1. Verified Properties

We described 2047 PSL properties in total and all properties were verified successfully as shown in Table 2. The table shows the module names, the number of sub-modules, the number of logic bugs found by formal verification, and the number of PSL properties for each target category. Note that the number of properties in Table 2 is only for assertions. It takes about 20 hours to verify all the properties on a typical Linux workstation with single CPU and single license.

The break down of the properties in number is as follows; 1306 properties are for checking ability of error detection, 200 properties are for checking soundness of internal states, 520 properties are for checking output data integrity, and 21 properties are for checking other features. We found 7 logic bugs in formal verification. This is about 5% of total logic bugs. The following sections describe the detail of each logic bug, the design impact, and the side effect of the error injection implementation.

### 6.2. Logic Bugs Found by Formal Verification

This section describes the detail of each logic bug. We found 7 logic bugs in formal verification. Table 3 shows the type of property and difficulty of finding the logic bugs by logic simulation.

After the detail analysis of results, B1, B3, B5 and B6 were turned out to be difficult to detect by logic simulation. The details are shown below.

**B1)** When a non-zero-value is written into a reserved field of a register, the internal parity of the register is not maintained correctly. Consequently, it causes an internal parity error. It is very difficult to find this type of logic bug by logic simulation because the scenario to hit this condition is very complicated.

**B3)** Although a signal comes from a macro is not guaranteed immediately after release of reset, the logic assumes a certain value with the signal. Therefore, a false parity error is logged and reported. The logic bug was not found by logic simulation because the behavior model of the macro cell was wrong. Actually, this is a problem of logic simulation environment, so it is impossible to find it by logic simulation.

**B5, B6)** An address decoder has ninety-one valid cases when decoding 8-bit address spaces. Parity calculation in the data path is wrong specifically for two cases out of ninety-one cases. The parity error is not always detected because it depends on the data pattern, so it is very difficult to detect it by logic simulation since it needs exhaustive data pattern.

In summary, at least four of seven logic bugs are difficult to detect by logic simulation, whereas they can be easily found by formal verification.

**Table 2. Number of verified properties**

| Module Name | # of Sub | # of Bug | Type of Property | | | | |
|---|---|---|---|---|---|---|---|
| | | | P0 | P1 | P2 | P3 | Total |
| A | 19 | 3 | 204 | 23 | 113 | 15 | 355 |
| B | 2 | 0 | 25 | 23 | 82 | 0 | 130 |
| C | 13 | 1 | 43 | 20 | 38 | 0 | 101 |
| D | 3 | 1 | 70 | 46 | 137 | 6 | 259 |
| E | 58 | 2 | 964 | 88 | 150 | 0 | 1202 |
| Total | 95 | 7 | 1306 | 200 | 520 | 21 | 2047 |

P0: Ability of Error Detection
P1: Soundness of Internal States
P2: Output Data Integrity
P3: Other Properties

**Table 3. Classification of logic bugs**

| Defect ID | Type of Property | Can be found by logic simulation easily? |
|---|---|---|
| B0 | Soundness of Internal States | Yes |
| B1 | Soundness of Internal States | No |
| B2 | Soundness of Internal States | Yes |
| B3 | Ability of Error Detection | No |
| B4 | Output Data Integrity | Yes |
| B5 | Output Data Integrity | No |
| B6 | Output Data Integrity | No |

### 6.3. Design Impact and Side Effect

This section describes the design impact and the side effect of the error injection implementation. Since the logic for the error injection remains in the netlist as shown in Figure 1, we checked the design impact in terms of area increase and timing delay using several modules. Basically, the design impact is caused by a selector added shown in Figure 1.

Table 4 shows the area increase caused by implementing the error injection feature, and the area increase is less than 2%. Note that module name in Table 4 corresponds to Table 2.

**Table 4. Area increase caused by implementing the error injection feature**

| Module Name | Area Increase |
|---|---|
| A | 1.4 % |
| B | 0.4 % |
| D | 0.2 % |

The timing delay of the selector is about 200 ps that are about 4 % of total delay when frequency is 250MHz. This timing delay was acceptable for the target chip and caused no timing closure issue.

In summary, the penalty caused by implementing the error injection feature is almost negligible in terms of area increase and timing delay.

One unexpected and good side effect is that those remaining gates can be used as spare logic gates. We performed ECO (post-route fixes) six times and we used these remaining gates twice.

## 7. Conclusions

We have applied formal verification method when we developed the component chip for servers with rich RAS feature. There are three reasons why we adopted formal verification.

- The target chip requires main-frame class reliability so that there are huge check points for data integrity. It is difficult to verify the data-centric properties exhaustively and effectively by conventional logic simulation.
- The description of the properties for integrity check is easy and simple because the properties are broken down into module level.
- The methodology above is clear for logic designers to release Verifiable RTL

Four logic designers and one verification engineer have developed more than 2000 PSL properties. Since the properties are written for leaf modules, it was easy to develop properties rather than logic simulation patterns.

As a result, we found seven logic bugs. Four of seven logic bugs were difficult to detect by logic simulation. Thus, it is shown that formal verification is powerful and effective.

We also investigated the design impact of implementing the error injection feature, and we found that the penalty was almost negligible in terms of area increase and timing delay.

What was novel in our approach was to break down properties for RAS features into three stereotype properties. Such framework enabled us to deploy formal methodology systematically and thoroughly over all the leaf modules. Each designer designed Verifiable RTL and described properties quite easily, although it was a very difficult task to complete verification for huge number of integrity checkpoints. As the result of our methodology, we conclude that significant increase of productivity is achieved.

## 8. Acknowledgement

We thank Takashi Miyoshi, Yoichi Koyanagi, Akihiko Okutsu, and Takuya Saze for their contribution of developing properties. We also thank Jawahar Jain and Christian Stangier for their support of using the in-house formal verification engine.

## 9. References

[1] K. L. McMillan. Fitting Formal Methods into the Design Cycle. In *Proceedings of DAC*, 1994

[2] R. P. Kurshan. Formal Verification In a Commercial Setting. In *Proceedings of DAC*, pp 258-262, 1997

[3] D.L. Dill. What's Between Simulation and Formal Verification ? In *Proceedings of DAC*, 1998

[4] D.L. Dill. Embedded tutorial: formal verification meets simulation. *In Proceedings of ICCAD*, 1999

[5] P.H.Ho, et al. Smart Simulation Using Collaborative Formal and Simulation Engines. In *Proceedings of ICCAD*, pp 120-126, 2000

[6] Tom Schubert. High Level Formal Verification of Next-Generation Microprocessors. In *Proceedings of DAC*, pp 1-6, 2003

[7] Maher Mneimneh, et al. Scalable Hybrid Verification of Complex Microprocessors. In *Proceedings of DAC*, pp 41-46, 2001

[8] Xi Chen, et al. Utilizing Formal Assertions for System Design of Network Processors. In *Proceedings of DATE-04*, pp126-131, 2004.

[9] PSL Language Reference Manual, version 1.01, 2003 http://www.eda.org/vfv/docs/psl_lrm-1.01.pdf

[10] J. Jain, Breaking Barriers of BDD-based Verification by Partitioning, IWLS 2004